# Vanadium oxide thin films and fibers obtained by acetylacetonate sol-gel method

O. Berezina[1], D. Kirienko[1], A. Pergament[1,*], G. Stefanovich[1], A. Velichko[1], V. Zlomanov[2]

[1] Department of Physical Engineering, Petrozavodsk State University, 185910, Petrozavodsk, Russia
[2] Department of Chemistry, Moscow State University, 119991, Moscow, Russia
* Corresponding author, email: aperg@psu.karelia.ru

**Abstract.** Vanadium oxide films and fibers have been fabricated by the acetylacetonate sol-gel method followed by annealing in wet nitrogen. The samples are characterized by X-ray diffraction and electrical conductivity measurements. The effects of a sol aging, the precursor decomposition and the gas atmosphere composition on the annealing process, structure and properties of the films are discussed. The two-stage temperature regime of annealing of amorphous films in wet nitrogen for formation of the well crystallized $VO_2$ phase is chosen: 1) 25-550°C and 2) 550-600°C. The obtained films demonstrate the metal-insulator transition and electrical switching. Also, the effect of the polyvinylpyrrolidone additive concentration and electrospinning parameters on qualitative (absence of defects and gel drops) and quantitative (length and diameter) characteristics of vanadium oxide fibers is studied.

## 1. Introduction

Due to the existence of an unfilled *d*-shell, vanadium possesses a set of valence states and, in compounds with oxygen, forms a number of oxides, namely, VO, $V_2O_3$, $VO_2$, Magneli ($V_nO_{2n-1}$) and Wadsley ($V_{2n}O_{5n-2}$) homologous series, as well as $V_2O_5$ [1, 2]. Many of these oxides are of considerable interest because of their potential applications. For example, vanadium dioxide exhibits a metal-insulator transition (MIT) at the transition temperature $T_t$ = 340 K accompanied by a conductivity jump of up to 4-5 orders of magnitude which can be utilized in thermochromic coatings, bolometric sensors, thin-film transistors and switches, and in a number of other optical and electronic devices [2, 3]. Vanadium pentoxide also finds multiple applications, e.g., in gas sensors, reversible cathode materials for Li batteries, electrochromic devices, etc. [4].

It should be noted that vanadium oxides are supposed to be used in nanoscale integrated optoelectronic devices in the form of thin films and nanostructures. Considerable efforts have been made in producing these in the last few decades, however, when compared, for example, to bulk $VO_2$ single crystals, where the MIT is accompanied by a sharp conductivity change within a very narrow temperature range at $T = T_t$, thin films and nano-fibers typically show more spread out transitions and less significant changes in their physical properties, which is due to various factors affecting the transition, such as non-stoichiometry, defects, crystallite grain boundaries, structural disorder, and dimensional effects [5]. That is why the improvement of the conditions for fabrication of more and more perfect films continues to be a topical problem.

The sol–gel technique is currently regarded as one of the most promising LPD (liquid phase deposition) processes, particularly, for obtaining thin oxide films, vanadium oxide included [4]. Like all LPD processes for depositing films and coatings, the sol–gel technique is notable for its simplicity and cheapness and requires no sophisticated process equipment, which is an indubitable advantage in mass production. In addition, this technique makes it possible to deposit thin-film coatings onto large-area and complex-shape substrates at low temperatures, which, in turn, is essential for applications in flexible electronics and 3D integrated circuits.

There are different routes for the vanadium oxide films sol-gel preparation – the hydrolysis of alkoxides, the technique based on melt quenching [4, 6-9], the acetylacetonate chemical method [10] in which the $VO_2$ films are deposited onto Si substrates using $VO(acac)_2$ as a precursor. VO $(acac)_2$ offers certain advantages when compared with vanadium alkoxides because of its stability against precipitation and excessive hydrolysis. Also, in this precursor, the valence of vanadium is four, therefore a reduction $V^{5+} \rightarrow V^{4+}$ is not required as in the case of a vanadium alkoxides precursor in which the valence of vanadium is five. Lastly, it has a relatively low price, low toxicity and possesses a possibility of doping by various elements [10].

In the acetylacetonate chemical method [10] the precursor sol was used for thin film spin coating only after aging for a week. After deposition the films were annealed at 80°C for 20 minutes to drive off the solvent. The obtained precursor films were turned into $VO_2$ phase after heat annealing for 30 minutes in $N_2$ at a temperature above 550°C. After that the films were annealed in $N_2$ at 600°C for 5 minutes to further improve the crystallization. The processes details of starting solution aging and heat treatment (e.g. a needed pressure of oxygen) were not reported in the work [10]. Note also that vanadyl acetylacetonate (oxy-bis(2,4-pentandione)-vanadium, $C_{10}H_{14}O_5V$, $VO(acac)_2$) is widely used in vanadium dioxide preparation by means of other chemical methods, for instance, by hybrid CVD [11].

Although in the acetylacetonate sol-gel method, as noted above, the stage of reduction annealing for the $V_2O_5$-to-$VO_2$ transformation to occur is not required, a certain thermal treatment process is still necessary to ensure the organic residua removal and oxide crystallization. Therefore, in this work we report on a study of the effect of preparation conditions, with emphasis on the starting solution aging and finishing annealing procedure, on the structures and properties of vanadium oxide films fabricated by the acetylacetonate sol-gel method. Vanadium oxide fibers obtained by the same method combined with electrospinning are also studied and their functional electrical properties arising from the MIT are discussed.

## 2. Experimental procedures

Thin films of vanadium dioxide were prepared by an acetylacetonate sol–gel method including the following steps.

1. Synthesis of VO(acac)$_2$ via interaction of VOSO$_4$ with acetylacetone:
   VOSO$_4$ + 2C$_5$H$_8$O$_2$ + Na$_2$CO$_3$ → VO(C$_5$H$_7$O$_2$)$_2$ + Na$_2$SO$_4$ + H$_2$O + CO$_2$.
2. VO(acac)$_2$ purification by means of recrystallization from chloroform.
3. Preparation of the VO(acac)$_2$ solution in methanol with a concentration of 0.125 mol/liter.
4. Deposition of the VO(acac)$_2$-methanol sol films onto substrates by spin coating.
5. IR drying at 150°C to remove the dissolvent, stimulate the VO(acac)$_2$ film formation and its transformation into amorphous VO$_2$.
6. Annealing of the amorphous VO$_2$ films.

Step 1 was made according to [12]. In step 3, the starting solution of the VO(acac)$_2$ in methanol was green-blow and did not adhere the substrate, however, after aging for 20 hours, the resulting solution became red-brown and it adhered the Si-substrate which could be associated with the solution composition changes.

According to [13] the VO(acac)$_2$ can be oxidized by molecular oxygen in commercial methanol MeOH (0.04 % v/v of H$_2$O) according to the following reaction:

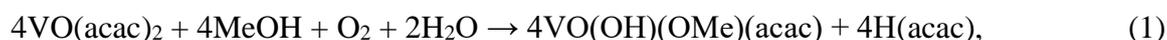

$$4VO(acac)_2 + 4MeOH + O_2 + 2H_2O \rightarrow 4VO(OH)(OMe)(acac) + 4H(acac), \qquad (1)$$

and the hydrolysis of VO(acac)$_2$ in acidic aqueous solutions proceeds in two stages, via reactions (1) and (2),

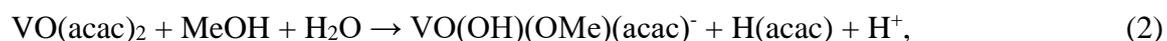

$$VO(acac)_2 + MeOH + H_2O \rightarrow VO(OH)(OMe)(acac)^- + H(acac) + H^+, \qquad (2)$$

with the liberation of the second acetylacetone molecule being 150 times slower than the first. Hydroxometoxooxo (pentane – 2.4 dionato) vanadium (V), written as VO(OH)(OMe)(acac) in reactions (1) and (2), leads to the red-brown solution color, and its "–OH" groups interact with "–Si-OH" syloxane groups of the Si-substrate and improve the adhesion at the deposition of the VO(acac)$_2$-methanol sol films onto the substrate.

The spinning procedure (step 4) started by first filling a syringe with 1 ml of sol and attaching a needle at the end of it. The spinning speed was 3000 rpm for 10-12 seconds. It should be mentioned that, at the first coating, the VO(acac)$_2$-methanol sol interacted with Si, Si-O and Si-OH groups of the Si substrate, since the silicon surface is covered with SiO$_2$ and syloxane Si-OH groups [14]. When the spinning run was completed the sample was heated (step 5), using an IR lamp to about $150^0$C in order to evaporate the remaining solvent and stimulate the VO(acac)$_2$ film formation. After the drying, the amorphous VO$_2$ layer was formed which was a substrate for the consequent coatings.

The process of deposition (steps 4 and 5) was thus carried out repeatedly to obtain a desired total film thickness (50 to 200 nm normally). Finally, the films were annealed in wet nitrogen under different temperature regimes, as will be discussed in Section 3.1 below, in order to optimize the procedure of crystalline vanadium dioxide phase formation.

Starting silicon substrates used were of p-type (Si:B, $\rho = 12$ $\Omega$·cm) and n-type (Si:P, $\rho = 50$ $\Omega$·cm). To afford electrical measurements, gold electrodes were deposited by thermal vacuum evaporation on the surfaces of vanadium oxide films through a mask, and the interelectrode gap was about 0.5 mm.

Vanadium oxide fibers were formed from a solution of VO(acac)$_2$ in methanol, with addition of polyvinylpyrrolidone (PVP) polymer, by an electrospinning method [15]. In this method, the sol flow was squirted out from a syringe with a 0.7 mm diameter needle with a constant flow rate in an external electric field and deposited onto a substrate. Thus, formation of microfibers occurred in the electrostatic field in the sol-polymer solution jet. Fluid supply from the syringe was carried out with a NE-300 pump, and aluminum foil was used as a collector for the fiber deposition. The voltage of 10 to 16 kV was applied between the needle and collector, and the tip-to-collector distance was varied in the range from 6 to 15 cm.

The solution feed rate from the syringe was varied between 0.05 and 0.50 ml/hr. When preparing the samples for electrical measurements, a glass-ceramics substrate, covered by pre-deposited gold contacts and an additional mask, was placed over the Al collector. The obtained fibers were then subjected to further drying and heat treatment.

The samples were characterized by X-ray diffraction (XRD) using a DRON-4 diffractometer (fibers) and a Rigaku D/MAX diffractometer (films), both with Cu-K$_\alpha$ radiation. The temperature dependences of resistance were measured by a four-probe technique using a Keithley 2410 SourceMeter supplied with a heater and copper-constantan thermocouple connected to a Keithley 2000 Multimeter. Current-voltage characteristics were measured with a Keithley 2636A SourceMeter.

## 3. Results and discussion

### 3.1. Vanadium dioxide films

The final 6th step of the film synthesis (Section 2) is the annealing of the obtained precursor films which would turn into crystalline $VO_2$ phase. During annealing the following processes can occur: sublimation of $VO(acac)_2$, decomposition of $VO(acac)_2$ accompanied by the formation of hydrogen, carbon monoxide, carbon dioxide, acetone, etc., the nucleation and growth of vanadium dioxide crystallites. To minimize the sublimation of the material, we have used the nitrogen atmosphere. Furthermore, the acetylacetone ligands and the organic residua can hinder the nuclei crystallization formation and their growth. Our preliminary results of the thermogravimetric (TGA) and differential thermal analysis (DTA), X-Ray and mass-spectroscopic measurements show that the $VO(acac)_2$ decomposition processes are completed below or around 500-550°C, while the crystallization is effective only above a temperature of 550°C. Thus, to lessen the mutual effect of the both processes, the two-stage temperature regime has been applied, namely: 1) 25-550°C and 2) 600°C (Table 1).

Also, when $VO(acac)_2$ decomposes, a reducing atmosphere containing CO and $H_2$ is formed, which can promote reduction of $VO_2$ to lower vanadium oxides. Besides, the $VO_2$ can be oxidized to $V_2O_5$ at a some oxygen partial pressures $P_{O(2)}$. The stable existence of the vanadium oxides depends on the temperature $T$, the $P_{O(2)}$ values and is determined by the $P_{O(2)}$–$T$ diagram [16, 17]. At $T = 600$°C the $VO_2$ oxide exists for example at $P_{O(2)} < 10^{-23}$ atm., and at $P_{O(2)} > 10$ atm. it is oxidized into $V_6O_{13}$ phase. A partial pressure $P_{O(2)}$ can be established by making use a mixture of gaseous compounds which tends to dissociate. Thus $H_2O$ and mixtures of $CO + CO_2$, $CO + H_2$ can be used to establish a well-defined oxygen pressure.

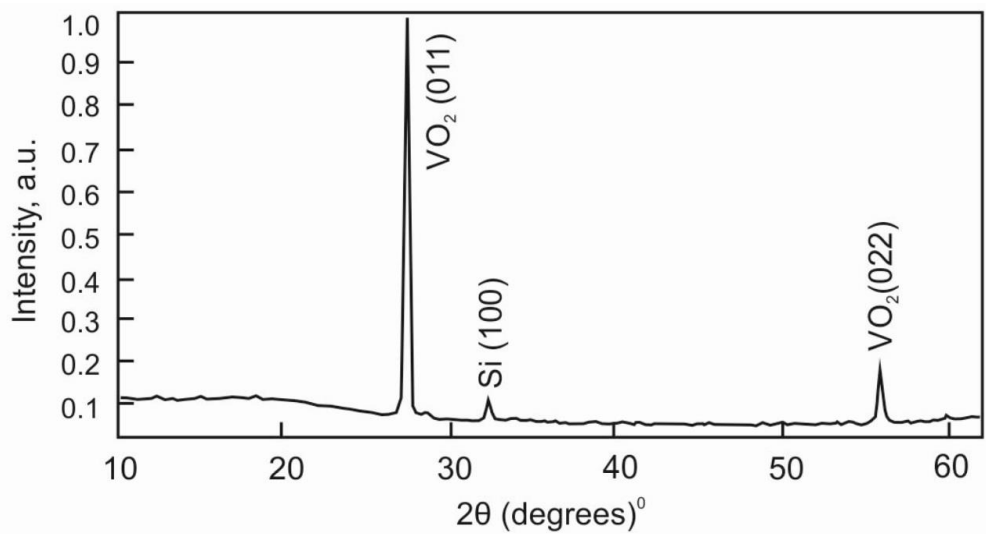

**Fig. 1.** XRD pattern of a vanadium oxide film annealed at regime IV (Table 1).

Optimal annealing regime is denoted as IV in Table 1. In the XRD spectra (Fig. 1) of the film prepared at this regime, there are two narrow peaks for the (011) and (022) planes, indicating the (011) orientation structure of $VO_2$. Besides, there is the peak at about $2\Theta = 32°$. It corresponds to the Si substrate that can be connected with the small film thickness (<100nm). Atomic-force microscopy reveals that the film surface is quite smooth and free from large granules (Fig. 2), and the temperature dependence of resistance for this sample (Fig. 3) demonstrates the conductivity jump at the transition of up to more than two orders of magnitude which is in agreement with the results of some other authors for vanadium dioxide thin films on silicon substrates (see, e.g., [8] and references therein).

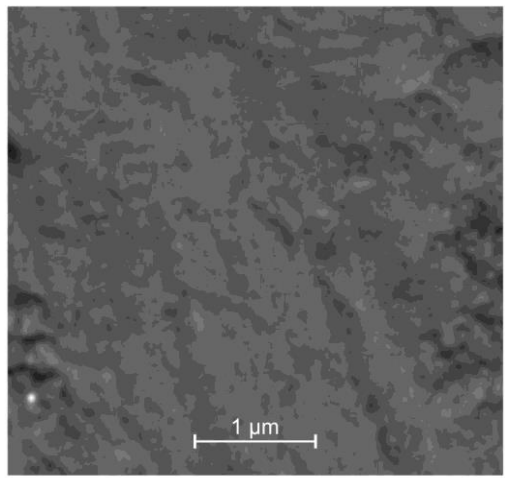

**Fig. 2.** Vanadium oxide film surface AFM image.

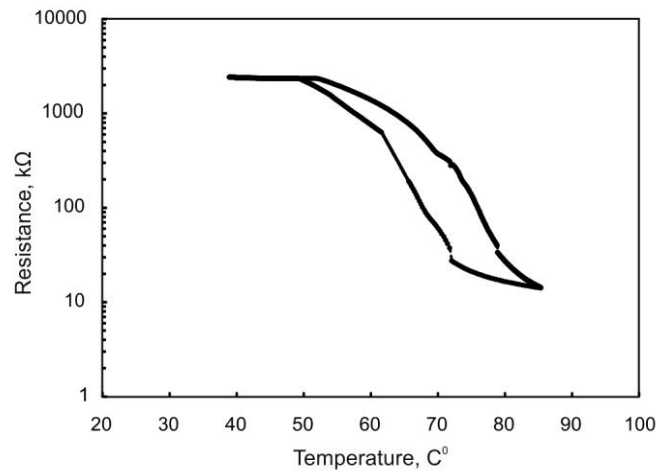

**Fig. 3.** Temperature dependence of resistance of the sample prepared with annealing regime IV. Film thickness is $d = 80$ nm.

To prevent the formation of lower vanadium oxides the process of annealing (step 6) was performed in the flow of wet nitrogen (water source was kept at 25°C, flowing rate 4-9 ml/min).

Thermal dissociation of water created a needed pressure of oxygen [18]. After annealing, the oven was switched off and the sample was cooled down to room temperature in a continued wet nitrogen flow to avoid a partial reduction. Quenching is not allowed since the film is destroyed due to the extreme temperatures difference. The parameters of the annealing conditions are summarized in Table 1.

Table 1. Annealing parameters.

|  |  | I | II | III | IV |
|---|---|---|---|---|---|
| Temperature regime | 1 | 25°C → 550 °C 15 min | 25 °C → 550 °C 15 min | 25 °C → 550 °C 15 min | 25 °C → 550°C 15 min |
| | 2 | 550 °C 30 min | 550 °C 30 min | 550 °C 35 min | 550°C 35 min |
| | 3 | 550 °C → 600°C 5 min | 550 °C → 600°C 5 min | 550 °C → 600°C 5 min | 550°C → 600°C 5 min |
| | 4 | 600 °C 15 min | 600 °C 5 min | – | – |
| Atmosphere | | Wet $N_2$, flowing rate 3-4 ml/min | The same | The same | Wet $N_2$, flowing rate 4-9 ml/min |
| Result (according to XRD) | | Traces of purely crystallized $VO_2$, significant amount of $V_2O_3$ | Moderately crystallized $VO_2$, some amount of $V_2O_3$ | Moderately crystallized $VO_2$, some amount of $V_2O_3$ | Well crystallized $VO_2$ (Fig.1) |

Next we measured the *I-V* characteristics of the films in a sandwich (Si-$VO_2$-Au) geometry and the results are presented in Fig. 4. After the process of electrical forming (EF), the switching effect due to an electronically-induced MIT in vanadium dioxide [19] is observed.

It should be emphasized that, unlike in MOM structures based on $V_2O_5$-gel films [4, 7] where EF is associated with the formation of a $VO_2$ channel inside the $V_2O_5 \times nH_2O$ matrix, in this case the film phase composition originally corresponds to vanadium dioxide and EF represents in fact the electrical breakdown of a thin silicon oxide layer that is likely to appear at the $VO_2$-Si interface during the thermal treatment.

The switching effect is conditioned by the development of an electrothermal instability in the $VO_2$ channel, arising between this breakdown channel, i.e. a Si pinhole in the $SiO_x$ film, and the top Au electrode. When a voltage is applied, the $VO_2$ channel is heated up to $T = T_t$ at a certain threshold voltage $V = V_{th}$ (e.g., $V_{th} \approx 7$ V for the negative branch in Fig. 4, a) and the structure undergoes a transition from an OFF insulating state to an ON metallic state. In high electric fields (~ $10^6$ V/cm), non-thermal electronic effects contribute to the Mott MIT in $VO_2$ modifying thereby the switching mechanism [3, 19]. Note that the current jump at this electrically driven transition (about ten times, Fig. 4) is lower than the conductivity jump at the thermally driven transition (more than 100 times, Fig. 3) due to a partial voltage drop across the Si substrate whose resistance is connected in series with the resistance of the vanadium dioxide channel [20].

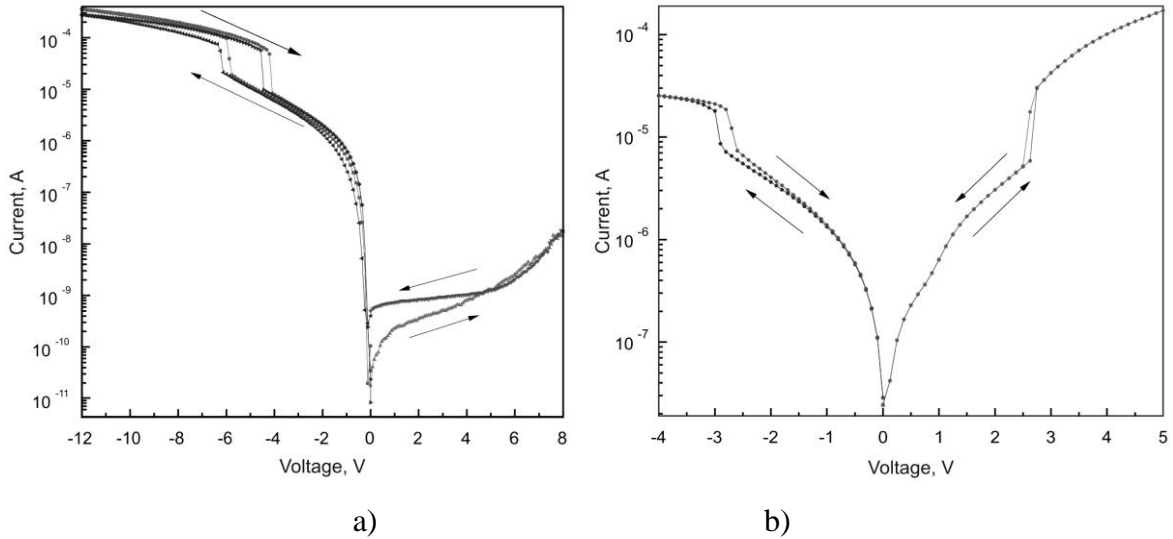

a) b)

**Fig. 4.** *I-V* curves of the $VO_2$ films on p-Si, $d = 175$ nm (a) and n-Si, $d = 190$ nm (b) [20].

We also note both qualitative and quantitative differences between the *I-V* curves of the samples on p-Si and n-Si (Figures 4 (a) and (b), respectively) which are supposed to be associated with the formation of isotype and anisotype heterojunctions and ion electrodiffusion processes at the Si/$VO_2$ interfaces. These effects are discussed in more detail elsewhere [20].

### 3.2. Vanadium oxide fibers

First we have studied the effect of the PVP concentration and the electrospinning parameters on qualitative (absence of defects and gel drops) and quantitative (length and diameter) characteristics of fibers.

At low polymer concentrations (region 1-2 in Fig. 5), vanadium oxide fibers are either not produced at all or produced with large defects in the form of droplets. Variation of the potential difference and the distance between the needle tip and the substrate does not reduce the number of defects in fibers. A further increase in the polymer concentration (region 3-5 in Fig. 5) results in the reduced number of defects and the improved quality of the synthesized microfibers. Also, the average fiber length increases with increasing the PVP concentration. The fiber images obtained using an optical microscope are shown in Fig. 6.

After the process of electrospinning and fiber synthesis, the fibers are annealed to remove the polymer and form crystalline vanadium oxide. Annealing is performed in a muffle furnace in an atmosphere of wet nitrogen. After the heat treatment the minimum fiber diameter ranges from 80 to 600 nm and depends on the annealing temperature and duration. XRD of the fibers annealed at 500°C for 60 minutes (Fig. 7, a) reveals complete removal of PVP and formation of the $V_2O_5$ phase.

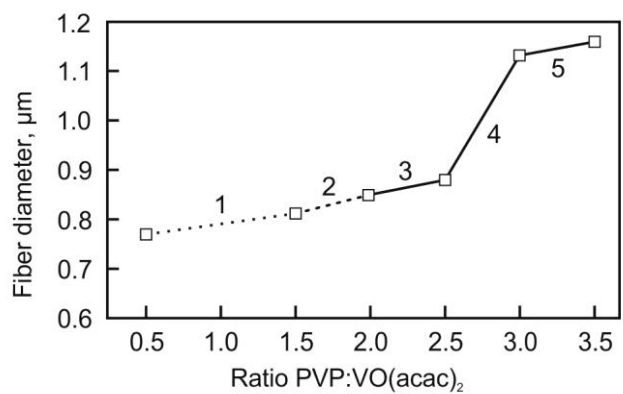

**Fig. 5.** Effect of PVP/VO(acac)$_2$ solution ratio upon the maximum diameter of the synthesized fibers.

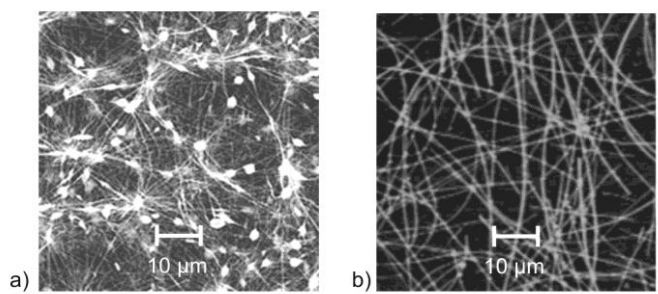

**Fig. 6.** Microscope images of vanadium oxide fibers obtained at the PVP:VO(acac)$_2$ ratios: a) 1.5:1 (droplet-like defects are seen) and b) 3:1.

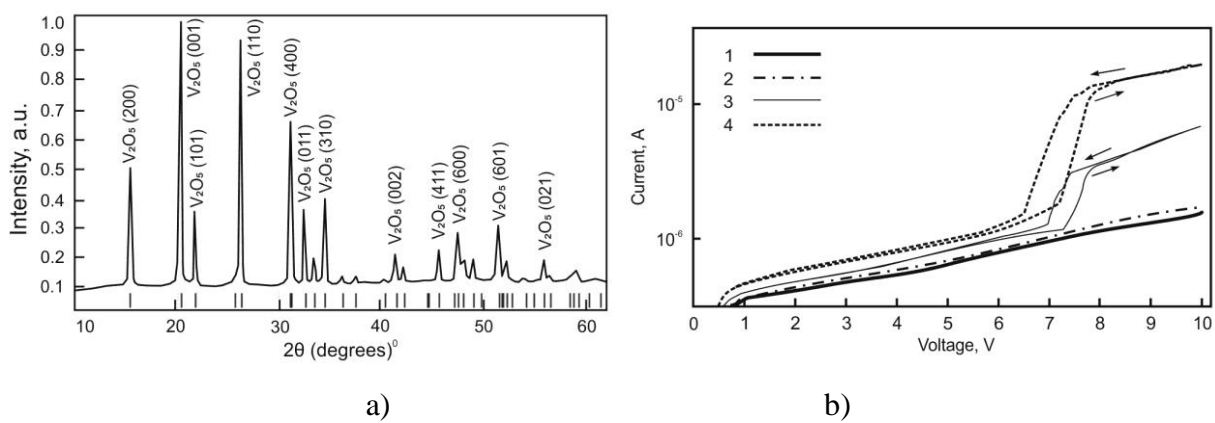

**Fig. 7.** (a) XRD pattern of fibers obtained at PVP:VO(acac)$_2$ = 2:1 and annealed at $T = 500$°C for 1 hour and Bragg bar chart for $V_2O_5$ with lattice parameters $a = 11.48$ Å, $b = 4.36$ Å, and $c = 3.55$ Å. (b) *I-V* characteristics of the fibers (PVP:VO(acac)$_2$ = 3:1) on glass-ceramic substrate with Au electrodes. The fibers were previously annealed at different regimes: 1 – $T = 400$°C, $t = 40$ min., 2 – 450°C, 40 min., 3 – 450°C, 60 min., and 4 – 500°C, 60 min. plus heating up to 600°C for 10 min.

The investigation of the electrical properties of vanadium oxide microfibers shows the occurrence of the switching effect (Fig. 7 (b), *I-V* curves 3 and 4) due to the MIT in vanadium dioxide. The fact that switching is associated with the MIT in $VO_2$ is supported by the following test: as the temperature increases, the threshold voltage decreases tending to zero at $T \sim T_t$ (~ 340 K). This experiment confirms that annealing regimes III and IV (Table 1) provide a sufficient amount of the vanadium dioxide phase, and this is the case not only for the films (Section 3.1), but for the fibers as well (Fig. 7 (b), curves 3 and 4).

## 4. Conclusion

Summarizing, in this work, vanadium oxide films and fibers have been successfully fabricated by the acetylacetonate sol-gel method followed by annealing in wet nitrogen. The effects of a starting sol aging, the precursor decomposition and the gas atmosphere composition on the annealing process, structure and properties of the films are discussed. The two-stage temperature regime of annealing of amorphous films in wet nitrogen for formation of the well crystallized $VO_2$ phase is chosen: warming-up at $550^\circ C$ for 35 min. with subsequent heating up to $600^\circ C$ during 5 minutes. The obtained films demonstrate the metal-insulator transition and electrical switching.

It should be admitted that the required thermal treatment temperature, of up to $600^\circ C$, is apparently rather high which would seem to eliminate such a major merit of the LPD technique as a low temperature processing temperature. Note however that even a relatively low content of the $VO_2$ phase (as, e.g., with conditions I or II in Table 1) would be enough to fabricate switching devices (Fig. 4), since $VO_2$ channels are always formed during the process of EF [4, 5, 7, 19, 20].

Also, the effect of the PVP concentration and electrospinning parameters on the qualitative (absence of defects and gel drops) and quantitative (length and diameter) characteristics of vanadium oxide fibers has been studied. It is found that both $V_2O_5$ and $VO_2$ fibers might be obtained by the above method.

Finally, the proposed procedures for preparation of vanadium oxide thin films and quasi-1D fiber structures based on the vanadyl acetylacetonate sol-gel route are promising in terms of developing functional devices of oxide nanoelectronics [3, 21].

**Acknowledgements**

This work was supported by the Strategic Development Program of Petrozavodsk State University (2012 – 2016), the RF Ministry of Education and Science as a base part of state program no. 2014/154 in the scientific field, projects no.1704 and no.3.757.2014/K, as well as through an RFBR grant no.13-03-00662.